\documentclass[aps,prd,onecolumn,superscriptaddress,showpacs]{revtex4}

\usepackage{graphicx}
\usepackage{times}
\usepackage{dcolumn}
\usepackage{color}

\newcommand{\be}{\begin{equation}}
\newcommand{\ee}{\end{equation}}
\newcommand{\bea}{\begin{eqnarray}}
\newcommand{\eea}{\end{eqnarray}}

\newcommand{\bx}{{\bf x}}
\newcommand{\bk}{{\bf k}}
\newcommand{\bq}{{\bf q}}
\newcommand{\bL}{{\bf L}}
\newcommand{\bP}{{\bf\Psi}}
\newcommand{\bs}{{\bf s}}
\newcommand{\cs}{{\cal S}}

\newcommand{\deltar}{\delta_{\rm recon}}

\begin{document}

\title{Reconstructing Baryon Oscillations : A Lagrangian Theory Perspective}

\author{Nikhil Padmanabhan}
\email{NPadmanabhan@lbl.gov}
\affiliation{Physics Division, Lawrence Berkeley National Laboratory,
1 Cyclotron Rd., Berkeley, CA 94720}

\author{Martin White}
\email{mwhite@berkeley.edu}
\affiliation{Departments of Physics and Astronomy, 601 Campbell Hall,
University of California Berkeley, CA 94720}

\author{J.D.Cohn}
\email{jcohn@berkeley.edu}
\affiliation{Space Sciences Laboratory, 601 Campbell Hall,
University of California, Berkeley, CA, 94720}

\date{\today}

\begin{abstract}
Recently Eisenstein and collaborators introduced a method to `reconstruct'
the linear power spectrum from a non-linearly evolved galaxy distribution
in order to improve precision in measurements of baryon acoustic oscillations.
We reformulate this method within the Lagrangian picture of structure formation,
to better understand what such a method does, and what the resulting power spectra are.
We show that reconstruction does {\it not\/} reproduce the linear density field,
at second order. We however show that it does reduce the damping of the oscillations
due to non-linear structure formation, explaining the improvements seen
in simulations. Our results suggest that the reconstructed power spectrum 
is potentially better modeled as the sum of three different power spectra, each
dominating over different wavelength ranges and with different non-linear damping terms.
Finally, we also show that reconstruction reduces the mode-coupling term in the 
power spectrum, explaining why mis-calibrations of the acoustic scale are
reduced when one considers the reconstructed power spectrum.
\end{abstract}


\maketitle
\twocolumngrid

\section{Introduction}

The baryon acoustic oscillation (BAO) method \cite{EisReview} is
an integral part of current and next-generation dark energy experiments.
Oscillations in the baryon-photon fluid, frozen into
the matter distribution at decoupling,
provide a standard ruler to constrain the expansion of the Universe.
These sound waves imprint an almost harmonic series of peaks
in the power spectrum $P(k)$, corresponding to a feature in
the correlation function $\xi(r)$ at $\sim$100 Mpc, with width $\sim 10$\%
due to Silk damping
(see \cite{Eis98,MeiWhiPea99} for a detailed description
of the physics, and \cite{ESW} for a comparison of Fourier and
configuration space pictures).
While the early Universe physics is linear and well understood, the low
redshift observations are complicated by the non-linear evolution of 
matter (not to mention galaxy bias and redshift space distortions \cite{SSS},
but we will defer these to future work) which erases the oscillations
on small scales and shifts the peaks \cite{ESW,CroSco,Mat08a,Seo08}
\begin{equation}
  P_{\rm obs}(k) = e^{-k^2\Sigma^2/2} P_{\rm lin}(k) + P_{\rm mc}(k)  +
  \cdots
\label{eq:processed}
\end{equation}
by coupling individual $k$-modes which are at early times independent.
The exponential damping of the linear power spectrum (or equivalently the 
smoothing of the correlation function) reduces the contrast of the feature
and thereby the precision with which the size of ruler may be measured.
Neglect or incorrect modeling of the ``mode-coupling'' term $P_{\rm mc}$
may bias the resulting distance measurements.

In \cite{ESW} it was pointed out that much of the modification to the
power spectrum comes from large-scale modes, bulk flows and super-cluster
formation, in principle enabling their effects to be corrected.
Eisenstein et al~\cite{ES3} introduced a method for removing the non-linear degradation
of the acoustic signature, sharpening the feature in configuration space or
restoring/correcting the higher $k$ oscillations in Fourier space.
Given the ambitious nature of future experiments, there has been considerable
interest \cite{ES3,Huff,Wagner} in ``reconstruction'' schemes which remove the
effects of non-linearities, reducing the damping and mode coupling terms above. 

Since the method proposed in \cite{ES3} is an inherently non-linear mapping of
the observed density field, it is difficult to intuitively understand.
It is however easily formulated within the 
Lagrangian picture of structure formation, where the fundamental
quantity is the displacement of particles from their initial positions
(contrasted with the Eulerian picture where one tracks the evolution of the
density field at a fixed location).  Motivated by recent developments in
Lagrangian perturbation theory (LPT) \cite{Mat08a,Mat08b}, we discuss
reconstruction within the context of LPT, both to elucidate how it works
and to expose possible shortcomings. 
Although we use the method of \cite{ES3} for specificity, the lessons learned
have broader validity.

We proceed as follows : \S\ref{sec:recon} introduces the 
essential aspects of both LPT as well as reconstruction.
We then compute the reconstructed density field to second order, and
demonstrate that there are corrections to the linear density at this 
order.
\S\ref{sec:power} then explains why the BAO feature is enhanced 
in the reconstructed power spectrum. 
We conclude in \S\ref{sec:discuss}, highlighting potential avenues
for improvements.

\section{Reconstruction and the Density Field}
\label{sec:recon}

The Lagrangian description of structure formation \cite{Buc89,Mou91,Hiv95}
relates the current, or Eulerian, position of a mass element, $\bx$, to
its initial, or Lagrangian, position, $\bq$, through a displacement vector
field $\bP(\bq)$,
\begin{equation}
  \bx = \bq + \bP(\bq) \,.
\end{equation}
The displacements can be related to overdensities by \cite{TayHam96}
\begin{equation}
  \delta(\bx) = \int d^3q\ \delta^{(D)}(\bx-\bq-\bP)-1 \; .
\end{equation}
where $\delta^{(D)}$ is the 3D Dirac $\delta$ function, or in Fourier space by 
\begin{equation}
  \delta(\bk) = \int d^3q\ e^{-i \bk\cdot \bq}
  \left(e^{-i \bk\cdot \bP(\bq)} - 1\right) \,\,.
\label{eq:lptdensity}
\end{equation}
The displacements
evolve according to
\begin{equation}
  \frac{d^2\bP}{dt^2} + 2 H \frac{d \bP}{dt} =
  -\nabla_x \phi\left[\bq+\bP(\bq)\right] \,\,,
\end{equation}
where $\phi$ is the gravitational potential.
Analogous to Eulerian perturbation theory, LPT expands the displacement in
powers of the linear density field, $\delta_l$,
\begin{equation}
  \bP = \bP^{(1)} + \bP^{(2)} + \cdots \; ,
\label{eq:psiexp}
\end{equation}
where \cite{BouColHivJus95}
\begin{eqnarray}
  \bP^{(n)}(\bk) &=& \frac{i}{n!} \int
  \prod_{i=1}^n \left[\frac{d^3k_i}{(2\pi)^3} \right] \nonumber \\
  &\times& \ (2\pi)^3\delta^{(D)}\left(\sum_i \bk_i-\bk\right) \nonumber \\
  &\times& \bL^{(n)}(\bk_1,\cdots,\bk_n,\bk)
  \delta_l(\bk_1)\cdots\delta_l(\bk_n) \; .
\end{eqnarray}
and the $\bL^{(n)}$ have closed form expressions, generated by
recurrence relations.  Specifically,
\begin{equation}
  \bL^{(1)} = \frac{\bk}{k^2}
\end{equation}
is the well known Zel'dovich displacement \cite[e.g.][]{GW87}, which is $1^{\rm st}$ order LPT.
Expanding the exponential in Eq.~(\ref{eq:lptdensity}) we obtain a
perturbative series for the overdensity,
$\delta = \delta^{(1)}+\delta^{(2)}+\cdots$
where, e.g.,
\begin{equation}
  \delta^{(2)} = \int d^3q\ e^{-i \bk\cdot \bq} \left[ -i \bk\bP^{(2)} - 
             \frac{(\bk\cdot\bP^{(1)})^{2}}{2}\right]
\label{eq:delta2}
\end{equation}
or in terms of the $L^{(n)}$'s
\begin{eqnarray}
\delta^{(2)} & = & \frac{1}{2} \int\frac{d^3k_1d^3k_2}{(2\pi)^3}
   \delta^{(D)}(\bk_1 + \bk_2 - \bk)  \nonumber \\
& \times &  \delta_{l}(\bk_1) \delta_{l}(\bk_2) \left[\bk \cdot \bL^{(2)}(\bk_1, \bk_2, \bk) \right. \nonumber \\
& + & \left. \bk \cdot \bL^{(1)}(\bk_1) \bk \cdot \bL^{(1)}(\bk_2) \right] 
\end{eqnarray}
is second order in the linear density field $\delta_l$.

The prescription of \cite{ES3} can be cast into this framework as follows:
\begin{itemize}    
\item Smooth the density field to filter out high $k$
non-linearities. In Fourier space, this is equivalent to 
multiplying by a function ${\cal S}(k)$ which monotonically decreases from
unity at low $k$ to zero at high $k$,
\begin{equation}
  \delta(\bk) \rightarrow {\cal S}(k) \delta(\bk) \; .
\end{equation}
\item Compute the negative Zel'dovich displacement from the smoothed density
field
\begin{equation}
  \bs(\bk) \equiv -i \frac{\bk}{k^2} {\cal S}(k)\delta(\bk) 
\end{equation}
\item Shift the original particles by $\bs$ and compute the ``displaced''
density field,
\begin{equation}
  \delta_d(\bk) = \int d^3 q e^{-i \bk\cdot \bq}
  \left(e^{-i \bk\cdot [\bP(\bq) + \bs(\bq)]} -1\right) \; . \\
\end{equation}
Note that if the original density field were linear, and $\mathcal{S}=1$,
this would undo their displacements exactly, moving them back to their
original positions and giving $\delta_d=0$.
\item Shift an spatially uniform grid of particles by $\bs$ to form
the ``shifted'' density field,
\begin{equation}
  \delta_s(\bk) = \int d^3 q e^{-i \bk\cdot \bq}
  \left(e^{-i \bk\cdot \bs(\bq)} -1 \right) \; , \\
\end{equation}
Again, assuming linear theory would imply $\delta_s(\bk)=-\delta(\bk)$.
\item The reconstructed density field is defined as
$\deltar \equiv \delta_{d} - \delta_{s}$
\begin{equation}
  \deltar(\bk) = \int d^3 q e^{-i \bk\cdot \bq} e^{-i \bk\cdot \bs(\bq)}
  \left(e^{-i \bk\cdot \bP(\bq)} -1\right) \; 
\label{eq:reconfield}
\end{equation}
with power spectrum
$P_{\rm recon}(k)\propto \langle \left| \delta_{\rm recon}^2\right|\rangle$.
\end{itemize}
Note that $\mathcal{S} \propto \bs \to 0$ is equivalent to no 
reconstruction, which is helpful in intepreting some of the expressions
below.

When applied to simulations this process yields an enhanced BAO feature
\cite{ES3,Huff,Wagner,Seo08} with a reduced `shift' in the peak.  Our
focus here is to understand what this procedure is doing within
an analytic framework.

In the spirit of LPT we can expand the reconstructed density field in a
perturbative series
\begin{equation}
  \deltar = \deltar^{(1)} + \deltar^{(2)} + \cdots \; .
\end{equation}
As anticipated above, the reconstructed field equals the linear density
field to lowest order.  Working to the next order, we find
\begin{eqnarray}
  \deltar^{(2)} &=& \delta^{(2)} - \frac{1}{2}\int\frac{d^3k_1d^3k_2}{(2\pi)^3}
  \delta^{(D)}\left(\bk_1+\bk_2-\bk\right) \nonumber \\
  &\times&\delta_l(\bk_1)\delta_l(\bk_2)
  \ \bk\cdot\bL^{(1)}(\bk_1) \bk\cdot\bL^{(1)}(\bk_2) \nonumber \\
  &\times& \left[ \mathcal{S}(\bk_1)+\mathcal{S}(\bk_2) \right] \; .
\label{eq:deltar2}
\end{eqnarray}
We observe that the second-order term in the reconstructed density field does
not vanish.  While $\delta^{(2)}$ contains $\bL^{(2)}$, the correction only
involves $\bL^{(1)}$ and so cannot fully cancel the non-linearity.
This is a general feature - the corrections to $\delta^{(n)}$ only involve
terms $\bL^{(i<n)}$ -- and follows from the fact that we only worked to first
order when shifting objects. We note in passing that one might be able to
construct higher order reconstruction schemes such that $\delta_{l}^{(n>1)}$
contributions to the reconstructed density vanish, but that is beyond
our scope here.

To recap: the reconstruction algorithm above generates a density field with 
second order corrections, {\it not\/} the linear density field.
The next section explains why simulations saw an improvement when using 
reconstruction, by considering the reconstructed power spectrum. 

\section{The Power Spectrum}
\label{sec:power}

\subsection{A Toy Model} \label{sec:toy}

To best highlight the effects of reconstruction on the power spectrum,
we start with a toy model that captures both the physics and the algebraic
structure of the full gravitational perturbation problem.  This toy model
is particularly useful for identifying the effect of reconstruction on
the nonlinear damping of the linear power spectrum in Eq.~\ref{eq:processed}.
\S~\ref{sec:LPT} describes the correspondence between the toy model and the full
gravitational instability problem, extending the analysis of
the effect of reconstruction
to the mode coupling terms as well.

Consider a model, inspired by the peak-background split, where $\bP$ can be
split into low ($L$) and high ($H$) frequency pieces,
\begin{equation}
  \bP = \bP_L + \bP_H \,,
\end{equation}
with $\bP_L$ the Zel'dovich displacement based on a linear density field
$\delta_l$, 
\begin{equation}
  \bP_L(\bk) = i \frac{\bk}{k^2} \delta_l(\bk) \,\,. 
\end{equation}
For simplicity we assume that $\bP_H$ is also Gaussian and is uncorrelated
with $\bP_L$.  The intuitive picture behind this model is that $\bP_L$ encodes
the linear density field, while $\bP_H$ encodes the non-linearities;
importantly, the baryon oscillations only exist in $\bP_L$ and not in $\bP_H$.

Using Eq.~(\ref{eq:lptdensity}) the power spectrum is
\begin{equation}
  P(k) = \int d^3 q e^{-i \bk\cdot \bq} 
  \left( \left\langle e^{-i k_{i} \Delta\Psi_{i}(\bq)} \right\rangle-1\right)
  \,,
\label{eq:lptpk}
\end{equation}
where $\bq = \bq_1 - \bq_2$, and $\Delta\bP = \bP(\bq_1) - \bP(\bq_2)$.
For Gaussian $\bP$
\begin{equation}
  \left\langle e^{-i \bk\cdot\Delta \bP(q)}\right\rangle =
  \exp \left[-\frac{1}{2}k_i k_j\left\langle 
  \Delta \Psi_i(\bq) \Delta \Psi_j(\bq) \right\rangle\right]
\end{equation}
with
\begin{equation}
k_i k_j\left\langle \Delta \Psi_{i}(\bq) \Delta \Psi_{j}(\bq) \right\rangle =
  2k_{i}^2\langle \Psi_i^2({\bf 0}) \rangle - 2 k_i k_j \xi_{ij}(\bq) \,\,,
\end{equation}
where $\xi_{ij}(\bq)\equiv\langle \Psi_{i}(\bq_1) \Psi_{j}(\bq_2)\rangle$
is the displacement correlation function and we have used translational
invariance for the correlation function at zero lag.
To lowest order the zero-lag correlation function is 
$\xi_{ij}({\bf 0})=\delta_{ij}\Sigma^2/2$, with $\Sigma^2$
the mean-squared Zel'dovich displacement of particles,
\begin{equation}
  \Sigma_L^2 = \frac{1}{3\pi^2} \int dp\ P_L(p)
\label{eq:sigmal}
\end{equation}
with a similar expression for $\Sigma_H$.
Note that the relation of the damping to the Zel'dovich displacement follows
naturally from the LPT formalism and shows the similarity of the treatments
in Refs.~\cite{CroSco,Mat08a,Mat08b} and \cite{ESW}.

\begin{figure}
\begin{center}
\leavevmode
\includegraphics[width=3.0in]{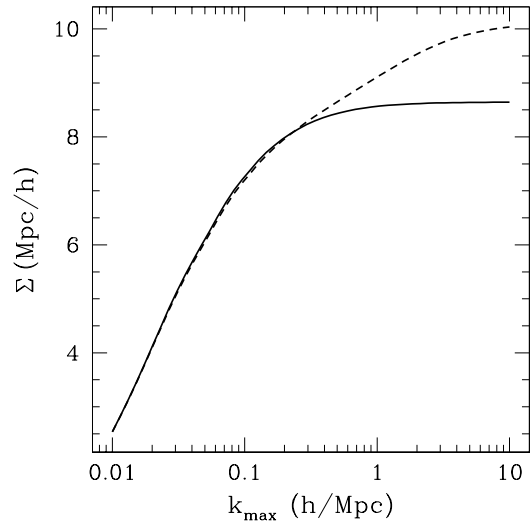}
\end{center}
\caption{The damping scale at $z=0$ as a function of the maximum wavenumber
for the linear (solid) and non-linear (dashed) power spectra.
Note that the dominant contribution to the damping scale come from linear
motions.}
\label{fig:variance}
\end{figure}

Given our assumption of uncorrelated low and high frequency pieces, we have
$\Sigma^{2} = \Sigma_{L}^2+\Sigma_H^2$.  However Fig.~\ref{fig:variance}
demonstrates that the dominant contribution comes from relatively large
($k<0.3\,h\,{\rm Mpc}^{-1}$) scales.
If we estimate $\Sigma_H$ by substituting the non-linear power spectrum in
the equation above, we find that the dominant contribution comes from linear
motions, even at $z=0$.
For simplicity, we will therefore assume $\Sigma^2\simeq\Sigma_L^2$ in what
follows.

The non-linear power spectrum is then given by 
\begin{equation}
  P(k) = e^{-k^2\Sigma_L^2/2} \int d^{3}q
  \ e^{-i k_i q_i}\ e^{k_i k_j \xi_{ij}(q)} \,.
\label{eq:toynl1}
\end{equation}
Following~\cite{Mat08a} we leave the zero-lag piece exponentiated, but expand
the exponential inside the integral.  The first term of the expansion gives
$P_L$.
This procedure can be viewed as a resummation of terms in the standard
perturbative expansion which leads to a power spectrum of the form in
Eq.~(\ref{eq:processed}),
\begin{equation}
  P_{\rm obs}(k) = e^{- k^2 \Sigma_{L}^2/2}
    P_L(k) + P_{\rm mc}(k) + \cdots
\label{eq:toynl2}
\end{equation}
with $P_{\rm lin}=P_L(k)$.
Note that $P_{\rm mc}(k)$ contains terms ${\cal O}(\bP_H^2)$ representing the
high frequency part of the power spectrum and terms $\mathcal{O}(\bP_L^4)$
corresponding to second order (in $P_L$) corrections.
We will consider these terms in the next section.

The above can be extended to compute the reconstructed power spectrum for this 
model. Since $\bP_H$ has no low frequency piece by construction, we assume that
the inferred shift, $\bs(\bk)$, is simply given by 
\begin{equation}
  \bs(\bk) = -\cs(k) \bP_{L}(\bk)  + O(\bP_L^2) \; .
\end{equation}
The fields $\delta_d$ and $\delta_s$ of Sec.~\ref{sec:recon} are then
generated to first order 
by $(1-\cs)\bP_L + \bP_H$ and $-\cs\bP_{L}$ respectively.
Since the reconstructed density field is the difference of the two fields,
there are three terms
(two auto-spectra, $P_{ss}, P_{dd}$ and one cross-spectrum $P_{sd}$) 
that make up the reconstructed power spectrum:
$P_{\rm recon} = P_{ss}+P_{dd} - 2P_{sd}$.
The auto-power spectra are exactly analogous to the non-linear power spectra,
except for the damping terms,
\begin{equation}
  P_{ss}(k) = e^{-k^2\Sigma_{ss}^2/2} \cs^2(k)P_L(k) + \cdots
\label{eq:pss_toy}
\end{equation}
and
\begin{equation}
\label{eq:pdd_toy}
  P_{dd}(k) = e^{-k^2\Sigma_{dd}^2/2} \left[1-\cs(k)\right]^2 P_L(k) + \cdots
\end{equation}
where we've dropped higher order terms.
The Gaussian damping is modified to
\begin{equation}
  \Sigma_{ss}^{2} =  \frac{1}{3\pi^2} \int dp\ \cs^2(p) P_{L}(p)
\label{eq:sigma_ss}
\end{equation}
with an analogous expression for $\Sigma_{dd}$ with $\cs^2 \rightarrow (1-\cs)^2$. The cross power spectrum is 
\begin{equation}
  P_{sd}(k) = -e^{-k^2\Sigma_{sd}^2/2} \cs(k) [1-\cs(k)] P_{L}(k) + \cdots
\label{eq:psd_toy}
\end{equation}
where
\begin{equation}
\Sigma_{sd}^{2} = \frac{1}{2} \left(\Sigma_{ss}^{2} + \Sigma_{dd}^{2}\right) \,\,,
\end{equation}
and the negative sign comes from the fact that the random field was shifted by
the negative Zel'dovich term.
Putting the pieces together, we find that the damping term becomes
\begin{eqnarray}
  D(k)\equiv e^{-k^2\Sigma^2/2} &\rightarrow &
  \cs^{2}(k) e^{-k^{2} \Sigma_{ss}^{2}/2} \nonumber \\
&+&  [1-\cs(k)]^{2} e^{-k^{2} \Sigma_{dd}^{2}/2} \nonumber \\ 
&+& 2 \cs(k) [1-\cs(k)] e^{-k^{2} \Sigma_{sd}^{2}/2}  \,\,.
\label{eq:damptransform}
\end{eqnarray}

\begin{figure}
\begin{center}
\leavevmode
\includegraphics[width=3.0in]{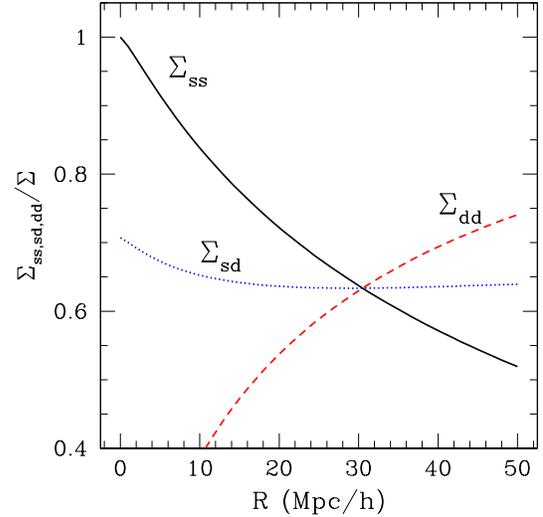}
\end{center}
\caption{The ratio of $\Sigma_{ss}$, $\Sigma_{sd}$ and $\Sigma_{dd}$ to
$\Sigma$, as a function of the Gaussian smoothing scale, $R$. Note that for
no smoothing, $\Sigma_{ss}=\Sigma$ and $\Sigma_{dd}=0$, while for infinite
smoothing, $\Sigma_{dd}=\Sigma$ with $\Sigma_{ss}=0$.}
\label{fig:sigmas}
\end{figure}

Before proceeding, it is useful to choose an explicit form for the smoothing;
the standard choice is a Gaussian, 
\begin{equation}
  \cs(k) = \exp \left(- \frac{k^2 R^2}{4}\right) \,\,.
\end{equation} 
Fig.~\ref{fig:sigmas} plots the various damping scales 
as a function of the smoothing scale. As expected, for non-zero smoothing, 
both $\Sigma_{ss}$ and $\Sigma_{dd}$ (and therefore $\Sigma_{sd}$ as well)
are less than the nonlinear damping scale.
This is the crux of the reconstruction method - that the $P_L$ contribution
to the reconstructed power spectrum is less damped than in the nonlinear
power spectrum.  This holds even when taking into account that there are
additional terms depending upon $\mathcal{S}(k)$ in $D(k)$ as we now show.

\begin{figure}
\begin{center}
\leavevmode
\includegraphics[width=3.0in]{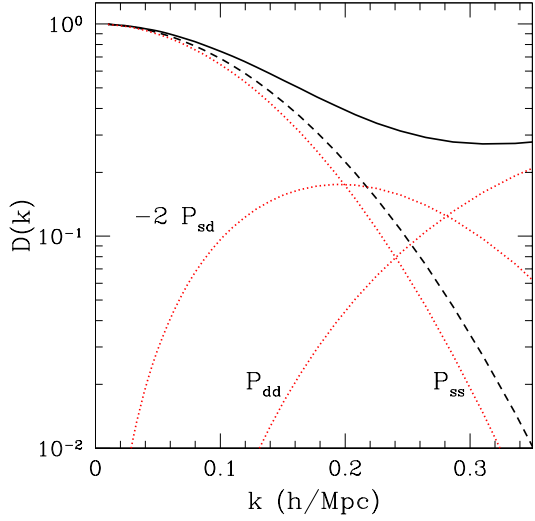}
\end{center}
\caption{The damping of the linear power spectrum for the nonlinear power
spectrum (dashed line), and the reconstructed power spectrum 
(Eq.~\protect\ref{eq:damptransform}, solid line, assuming a smoothing scale
$R=5\,h^{-1}$Mpc). 
The dotted lines decompose the reconstructed damping into the leading
contributions from its $P_{ss}$,
$P_{sd}$ and $P_{dd}$ components
These curves have been calculated assuming $z=0$.}
\label{fig:reconvar}
\end{figure}

Before considering Eq.~\ref{eq:damptransform} for arbitrary choices of
smoothing scales, we consider the special case where
$\Sigma_{ss}=\Sigma_{dd}=\Sigma_{sd}$;
for the Gaussian smoothing above, this corresponds to a smoothing scale
$R\sim 30\,h^{-1}$Mpc.  The damping of $P_L$ in the reconstructed power
spectrum simplifies considerably; the reconstructed power spectrum has
the form,
\begin{equation}
  P_{\rm recon}(k) = e^{-k^2\Sigma_{ss}^2/2} P_L(k) + \cdots
\end{equation}
Note that this is identical to the form of the nonlinear power spectrum
(Eq.~\ref{eq:toynl2}) except that $\Sigma_{ss}<\Sigma$, reducing the damping.

Fig.~\ref{fig:reconvar} shows the damping, $D(k)$, for smoothing scale
$R=5\,h^{-1}$Mpc, as
an example of its general form for an arbitrary choice of smoothing scale.
Given $R$, the factors involving $\cs(k)$ determine the range of wavenumbers
for which each of the three power spectra dominate.
For large $R$ (see below), $P_{dd}$ dominates over the wavenumbers important
for baryon oscillations ($0.07<k (h/{\rm Mpc})<0.35$) but as we argue below,
this limit is not optimal.  As we decrease $R$, we might have expected that
$P_{ss}$ would have dominated; however, decreasing $R$ quickly increases
$\Sigma_{ss}$ to close to the nonlinear damping scale (Fig.~\ref{fig:sigmas}),
limiting the importance of $P_{ss}$.
Indeed, in Fig.~\ref{fig:reconvar}, we see that $P_{ss}$ has the linear power
spectrum more strongly damped than the nonlinear power spectrum.
The dominant term at small $R$ is therefore $P_{sd}$; Fig.~\ref{fig:sigmas}
shows that $\Sigma_{sd}$ is $\sim 0.6\,\Sigma$ and is only weakly dependent
on $R$. This suggests that such a reconstruction method can reduce
the damping of the linear power spectrum by a factor $\sim 2$.

The above discussion argues that the smoothing scale determines the wavenumber
where $P_{dd}$ becomes dominant.  The obvious question is whether the above
analysis suggests a value for the smoothing scale.
We argue that the natural choice is $R\sim \Sigma$, the nonlinear (damping)
scale.  To see why, we start by observing that the terms we ignored in
$P_{dd}$ are ${\cal O}(\bP_H^2)$, whereas for $P_{ss}$ and $P_{dd}$ they
involve higher powers of the displacement.
This is just the statement that the small-scale displacements have their
largest effect on $P_{dd}$ which is not surprising, given that $P_{dd}$ is
based on the original density field.  We would ideally want to reduce these
terms, which argues for making $R$ as small as possible.
However, from Eq.~\ref{eq:toynl2}, we see that the linear field is damped
on scales smaller than $\Sigma$.  Smoothing on scales much smaller would
then violate our assumption that $\bs(\bk)$ is derived from the linear
density field, which leads to choosing $R\sim \Sigma$ as might have intuitively
been expected.
 
The above discussion explains how reconstruction reduces the damping of the
acoustic oscillations (or equivalently, how it sharpens the peak in the
correlation function).  We now turn to its effect on the mode-coupling terms,
by considering the reconstructed power spectrum within LPT.

\subsection{Lagrangian Perturbation Theory}
\label{sec:LPT}

Many of the features of reconstruction in the last section carry across to
the gravitational instability problem within LPT.  We will closely follow the
LPT formalism developed in \cite{Mat08a,Mat08b} in which the broadening of
the peak and the mode coupling terms appear naturally.

For the un-reconstructed power spectrum the derivation leading to
Eq.~(\ref{eq:lptpk}) still holds.  However now we must use the cumulant
expansion theorem
\begin{equation}
  \left\langle e^{-iX}\right\rangle =
  \exp\left[\sum_{N=1}\frac{(-i)^N}{N!}\left\langle X^N\right\rangle_c\right] 
\end{equation}
(where the $\langle X^N \rangle_{c}$ are the connected moments)
to compute the expectation value of the exponential.  In the toy model only
the $N=2$ term survived, for the full problem higher orders contribute as well.
Expanding $(\bk\cdot\Delta\bP)^N$ using the binomial theorem we have two
types of terms: those where the $\bP$ are all evaluated at the same point
(which we can take to be the origin) and those with $j$ $\bq_1$s and
$N-j$ $\bq_2$s.
As in the toy model, and following \cite{Mat08a}, we leave the first set of
terms exponentiated while expanding the second set of terms in powers of $\bP$.
If we keep only the lowest order terms in the exponential we regain the form of
Eq.~(\ref{eq:processed}) with $\Sigma$ given by the rms Zel'dovich displacement
\begin{eqnarray}
  P(k)  &=& e^{-k^2\Sigma^2/2}\left\{ P_L(k)\vphantom{\int} \right.\nonumber\\
  &\times& \left[1+\int d^3k_1\ P_L(k_1) G(\bk,\bk_1)\right] \nonumber \\
  &+& \int d^3k_1d^3k_2\ P_L(\bk_1)P_L(\bk_2)F^2(\bk_1,\bk_2,\bk) \nonumber\\
  &+& \left. \cdots \vphantom{\int} \right\}
\end{eqnarray}
where $F$ and $G$ can be expressed in terms of $\bL^{(1)}$ and $\bL^{(2)}$
and explicit expressions may be found in \cite{Mat08a}.
There are oscillations in $P_L$ and the mode-coupling term (third line), but
the integral in the second line has a wide kernel so the oscillations
are suppressed.
As it happens, $F$ is peaked around $\bk_1\approx\bk_2\approx\bk/2$, which
helps to explain why this the term in the third line leads to a peak shift.
If $P_L$ contains an oscillatory piece, e.g.~$\sin(kr)$, then the third term
contains a piece schematically of the form $\sin^2(kr/2)\sim 1+\cos(kr)$,
which oscillates out of phase with $P_L$.  It is the sum of the two out of
phase components that leads to a shift in the peak of $\xi(r)$ or the
phasing of the harmonics in $P_{\rm obs}(k)$.

It is now straightforward, though tedious, to repeat these steps for the
reconstructed field.  The formalism of Ref.~\cite{Mat08a} must be generalized
to allow two displacements ($\bs$ and $\bP$).
Again there are three contributions, $P_{ss}$, $P_{dd}$ and $P_{sd}$, and
three smoothings, $\Sigma_{ss}$, $\Sigma_{dd}$ and $\Sigma_{sd}$ of the same
form as before.  The term proportional to $P_L$ becomes
\begin{equation}
\begin{array}{ll}
  P_{\rm recon}(k) &= \left\{e^{-k^2\Sigma^2_{ss}/2} {\cal S}^2(k) \right. \\
  & + 2 e^{-k^2 \Sigma^2_{sd}/2}
        {\cal S}(k)\left[1-{\cal S}(k)\right] \\
  & + \left. e^{-k^2\Sigma^2_{dd}/2} \left[1-{\cal S}(k)\right]^2
      \right\} P_L(k) \, ,
\end{array}
\end{equation}
directly comparable to the result of the toy model.

The leading contribution to
the mode coupling term is the same as in standard perturbation theory,
and is strictly positive, coming from
$\left\langle \delta^{(2)}\delta^{(2)}\right\rangle$.
Recalling the relation between $\deltar^{(2)}$ and $\delta^{(2)}$ we need
to replace $F$ in the mode-coupling term with $\widehat{F}$ s.t.
\begin{eqnarray} 
  2\widehat{F}(\bk_1,\bk_2) &\equiv& \bk\cdot\bL^{(2)}(\bk_1,\bk_2)+
  \bk\cdot \bL^{(1)}(\bk_1) \bk\cdot \bL^{(1)}(\bk_2) \nonumber \\
  &\times& \left[1-\mathcal{S}(\bk_1)-\mathcal{S}(\bk_2)\right]
\end{eqnarray}
where $\bk_1+\bk_2=\bk$.  The piece of the mode-coupling integral which
shifts the peak comes from $\bk_1\approx\bk_2\approx\bk/2$ where
$\bk\cdot\bL_2=0$ and
\begin{equation}
  \widehat{F}(\bk/2,\bk/2) = 2\left\{1-2\mathcal{S}(\bk/2)\right\} \, .
\end{equation}
Since the term in $\{\cdots\}$ is bounded between $-1$ and $1$, 
$\hat{F}^{2} < F^{2}$ for all $k$, suppressing the reconstructed
mode-coupling term relative to the corresponding term in the non-linear
power spectrum. This explains why the mis-calibrations in the acoustic
scale were reduced after reconstruction in \cite{Seo08}. 

\section{Comments} \label{sec:discuss}

It is now generally understood (\cite{ESW, CroSco, Mat08a} and this work) 
that the dominant effect of the non-linear evolution of matter perturbations 
on the baryon oscillations is to damp the higher harmonics, $P_{\rm obs}(k) = 
\exp(-k^2 \Sigma^2/2) P_{\rm lin}(k) + \cdots$, or equivalently, smooth
the feature in the correlation function. Eisenstein et al~\cite{ES3} proposed a ``reconstruction'' 
method, demonstrated on simulations, that undoes this non-linear smoothing
and appears to restore the linear power spectrum. Motivated by recent
progress in Lagrangian perturbation theory \cite{Mat08a, Mat08b}, 
we revisit this algorithm in order to better understand why
it works as well as its shortcomings. Our principal conclusions are  
\begin{itemize}
\item[(i)] The field generated by the reconstruction process is {\it not\/}
the linear density field at second order. Note that this is a general statement,
independent of assumptions about the smoothing of the initial density field.
\item[(ii)] Reconstruction does reduce the damping of the oscillations, by
about a factor of 2 when the input density field is smoothed on the non-linear
scale.
\item[(iii)] Reconstruction also reduces the mode coupling terms which
introduce an out of phase component of the oscillations or shift the peak.
\item[(iv)] The reconstructed power spectrum is the sum of three power 
spectra (the auto-power spectra of the displaced and shifted fields, and their
cross-spectrum), each of which have different damping terms (Eq.~\ref{eq:damptransform}). 
An appropriate model for the reconstructed power spectrum should take this into account,
instead of modeling it as a single damping scale. 
\item[(v)] When the smoothing scale is close to the non-linear scale, the
correlation between the shifted and displaced fields plays a crucial role.
\end{itemize}

Our results suggest a number of natural extensions. The effects of bias and
redshift space distortions have been incorporated into the Lagrangian 
formalism \cite{Mat08a, Mat08b}, and could therefore be folded in to 
the LPT formulation of reconstruction. We have observed that the reconstructed 
density field is not the linear density field; an interesting possibility is
to explore whether higher order reconstruction schemes actually yield 
dividends. Even within the context of the existing reconstruction schemes, it 
is possible that a different weighting of the three power spectra may yield
improved accuracy in measuring the distance scale. We leave these avenues open 
for future investigation.

\acknowledgments

We thank David Spergel and Will Percival for conversations on reconstruction.
NP is supported by NASA 
NASA HST-HF-01200.01 and LBNL.
MW is supported by NASA and the Department of Energy.
JC is supported by the Department of Energy.
This work was supported by the Director, Office of Science, of the U.S. 
Department of Energy under Contract No. DE-AC02-05CH11231.


\bibliography{paper}
\bibliographystyle{apsrev}

\end{document}